\documentstyle[10pt,aaspp4]{article}

\begin{document}

\begin{center}
\title{A SHOCK-PATCHING CODE FOR ULTRA-RELATIVISTIC FLUID FLOWS}
\author{Linqing Wen\altaffilmark{1}, Alin Panaitescu \& Pablo Laguna\altaffilmark{2}}
\affil{Department of Astronomy \& Astrophysics,\\
       Pennsylvania State University, University Park, PA 16802}
\altaffiltext{1}{currently Physics Department, M.I.T., Cambridge, MA 02139, lqw@mit.edu}
\altaffiltext{2}{also Center for Gravitational Physics and Geometry,
                 Pennsylvania State University, plaguna@astro.psu.edu}
\end{center}

\bigskip

\begin{abstract}

We have developed a one-dimensional code to solve ultra-relativistic
hydrodynamic problems,
using the Glimm method for an accurate treatment of shocks and contact
discontinuities. The implementation of the Glimm method
is based on an exact Riemann solver and van der Corput sampling sequence. 
In order to improve computational efficiency, the Glimm method is replaced by
a finite differencing scheme in those regions 
where the fluid flow is sufficiently smooth.
The accuracy and convergence of this hybrid method is investigated 
in tests involving planar, cylindrically and spherically symmetric flows that exhibit 
strong shocks and Lorentz factors of up to $\sim 2000$. 
This hybrid code has proven to be successful in simulating
the interaction between a thin, ultra-relativistic, spherical shell and a low density 
stationary medium, a situation likely to arise in Gamma-Ray Bursters, supernovae
explosions, pulsar winds and AGNs.

\end{abstract}


\section{ Introduction}

One of the challenges in ultra-relativistic hydrodynamic problems for
compressible fluids is handling the sharp discontinuities at shocks and contact
discontinuities (CDs). There is a considerable number of situations in astrophysics
requiring accurate, shock capturing
methods for relativistic flows. Among them is the simulation of the interaction
of a cold shell expanding into a stationary 
external medium -- a popular model for Gamma-Ray Bursts 
(GRBs -- \cite{meszaros93}; \cite{laguna93}).
To explain the short burst durations
and observed fluxes, the blasting shell must be ultra-relativistic, with
Lorentz factors of up to a few thousands. 
As a result of the collision, strong shocks propagate in both, the expanding
shell and the external medium. 
The deceleration process is slow if the external 
medium has a density of $\sim 1$ particle/${\rm cm}^3$.
In this situation,
the evolution time of the shocked structure is $10^4-10^6$ times larger than its
crossing time, thus requiring a large computational effort. Numerical simulations
of the shell--external medium interaction at the lowest Lorentz factors ($\sim 100$) that may
generate the observed GRBs have been published by \cite{alin97}.

The main goal of this work is to construct a computationally efficient code based on an
algorithm that accurately resolves discontinuities during long term evolutions.
Finite difference (FD) methods with artificial viscosity have been 
a popular choice when dealing with shocks; however these methods
smear shocks and contact discontinuities unless implicit updates and/or adaptive-mesh refinements 
are used (Norman \& Winkler 1986, Woodward \& Colella 1984). 
The Piecewise Parabolic Method (PPM -- \cite{colella84})
has provided a powerful and accurate alternative to treating strong shocks, and 
has recently been generalized to relativistic flows (Mart\'{\i} \& M\"uller 1996).
We expect the PPM to be computationally too expensive for our problem,
due to the rather lengthy procedures it involves.
We chose to use the ``random choice" (or Glimm) method to develop a code that
simulates the shell--stationary medium interaction over a long time, due to
its computational efficiency and robustness in problems involving long
term evolutions of discontinuities.

The theoretical foundation of the random choice method is due to \cite{glimm65} and 
consists of two steps: 
(1) the fluid is approximated at each time-step by piecewise constant states,
and local Riemann problems formed by the neighboring states are solved; 
(2) the solution at the next time-step is taken to be the
exact solution of these Riemann problems at a  point randomly chosen in each cell.
\cite{chorin76} developed the Glimm method into a numerical one
for homogeneous hyperbolic conservation laws.
\cite{sod78}, in his survey of finite difference methods, studied the quality
of the Glimm scheme using a 1D shock tube problem. This method was
found to be first order accurate and to provide the best resolution of shocks and
CDs. 
However, as pointed out by Colella (1982), due to the random sampling used by Sod, 
the rarefaction waves were not completely smooth and the position 
of discontinuities were not very accurate, 
which led to an under-estimation of the overall quality of the Glimm method.
\cite{colella82} proposed a better procedure of randomly sampling the solution
of local Riemann problems and investigated the extension of the Glimm method
to two dimensions using the operator splitting method. 
He found that in 1D the Glimm method is superior to
any FD method when computing shock fronts,
in transporting discontinuities at the correct speed and in giving the correct
shape of continuous waves. 

The Glimm method has also been extended to inhomogeneous systems of conservation laws. 
For instance, \cite{sod77} used operator splitting to extend the Glimm method
to cylindrically symmetric flows. \cite{marshall81} solved non-conservative Riemann
problems by integrating along characteristics. 
\cite{liu79} introduced a generalized Glimm scheme to construct global
solutions for quasi-linear hyperbolic systems.
Based on Liu's work, \cite{glimm84} developed their generalized
random choice method: the solution at each time-step is approximated by a
piecewise steady flow and is advanced to the next time step by solving
a ``generalized" Riemann problem. 

Our work demonstrates that 
the Glimm method is appropriate for simulating 1D ultra-relativistic flows
involving strong shocks.
As mentioned before, the Glimm method requires solving a Riemann problem
for every two neighboring cells and thus could easily become expensive. 
Our approach to overcome this problem is to limit the use of
the Glimm method only to those regions of the fluid where steep gradients are
present and to apply a FD scheme in the smooth remaining part of
the computational domain. This approach is suggested by the fact that
the numerical complexity of the Glimm method
pays off mainly in the accurate representation and propagation of discontinuities.

\section{Numerical Techniques}

The special relativistic hydrodynamics (SRHD) equations 
governing the dynamics of 1D perfect fluids can be written as:
\begin{equation}
\frac{\partial \rho}{\partial t} = - v\frac{\partial \rho}{\partial r} -
\frac{1}{1-v^2 c_s^2} \left[ - \frac{v\rho}{\gamma^2 h}
\frac{\partial p}{\partial r} + \rho \frac{\partial v}{\partial r} +
\alpha \frac{v\rho}{r} \right] \, ,
\label{dddt}
\end{equation}
\begin{equation}
\frac{\partial p}{\partial t} = - \frac{1}{1-v^2 c_s^2} \left[ v(1-c_s^2)
\frac{\partial p}{\partial r} + \Gamma p
\frac{\partial v}{\partial r} + \alpha \frac{\Gamma v p}{r} \right] \, ,
\end{equation}
\begin{equation}
\frac{\partial v}{\partial t} = - \frac{1}{1-v^2 c_s^2} \left[
\frac{1}{\gamma^4 h}
\frac{\partial p}{\partial r} + v(1-c_s^2)
\frac{\partial v}{\partial r} - \alpha \frac{v^2 c_s^2}{\gamma^2 r} \right] \, ,
\label{dvdt}
\end{equation}
with $\rho$ the co-moving mass density, $p$ the pressure and $v$ the velocity 
of the fluid.
Above, $\gamma$ denotes the Lorentz factor of the flow,
$h=\rho+e+p$ the enthalpy density and
$c_s=\sqrt{\Gamma p/h}$ the local sound speed.
The equation of state for an ideal fluid $e=p/(\Gamma-1)$ is assumed, where $e$ is 
the internal energy density and 
$\Gamma$ the adiabatic index. The speed of light is set to 1.
The last term in the above equations is a geometrical term, with
$\alpha = 0$, 1, and 2 for planar, cylindrical and spherical symmetry, respectively.
One often finds in the literature the above SRHD
equations written in terms of quantities measured in a fixed ``laboratory" frame:
mass $D=\gamma\rho$, momentum $S =\gamma^2 v h$ and energy
$E =\gamma^2 h-p-D$ densities (e.g. Hawley, Smarr \& Wilson 1984). 
The use of $D$, $S$ and $E$ yields equations which explicitly exhibit a
conservation form similar to that of their non-relativistic counterparts.
The drawback of solving the SRHD equations in the conservative form  
is that iterations must be done in every grid zone
to determine the basic variables ($\rho,p,v$) from the set ($D,S,E$) whenever the
variables $(\rho,p,v)$ are required for physical applications, thus substantially
increasing the computational effort. Even more, the variables $(\rho,p,v)$ 
are required for an easier implementation of the most important ingredient of the Glimm method:
the Riemann solver (see below).

We describe first our numerical approach to solve in planar symmetry the problem
consisting of an initial discontinuity that separates two uniform states 
(Riemann problem). The solution to the Riemann problem 
in Newtonian hydrodynamics was derived by \cite{godunov59}. 
A detailed direct implementation
of Godunov's derivation to compute numerical solutions has been
given by \cite{chorin76} and \cite{sod78}. 
\cite{harten81} replaced the Riemann solver with a finite
difference approximation; \cite{roe81} constructed a local linearization of the Riemann
problem and \cite{colella82} represented the Riemann problem by two shocks. 
The analytic solution to the Riemann problem in relativistic hydrodynamics
was derived by \cite{thompson86}, in the particular case when the initial
states are at rest; the general case has been studied analytically by \cite{smoller93} 
and by \cite{marti94}. 

In any Riemann problem, Newtonian or relativistic, the initial discontinuity decomposes into
a contact discontinuity (CD) and two other elementary waves that can be
either a shock or a rarefaction wave.
The pressure and the flow velocity are constant between these elementary waves, 
with the density having a jump across the CD. The general features of
the solution of a Riemann problem can be seen in Figure 1. In all graphs, the density profile shows
from left to right: left initial state, rarefaction fan, left post-wave state, CD, right post-wave
state, shock, right initial state. All states, excepting the rarefaction fan, are uniform, i.e$.$ the
gradients of $\rho$, $p$ and $v$ are zero.
The waves are uniquely determined by the state of the pre-wave fluid (either left or right)
$(\rho,p,v)_{L/R}$ and the post-wave pressure $p_*$. The first step in solving a Riemann problem
is to determine the post-wave states, i.e$.$ the two uniform states 
($\rho_{*L},p_*,v_*$) and ($\rho_{*R},p_*,v_*$)
around the CD. We further list the equations that are used in the Riemann solver to
calculate the density $\rho_*$ and velocity $v_*$ behind each of the two elementary waves 
that develop in a Riemann problem, given the pre-wave state ($\rho_0,p_0,v_0$) and the pressure $p_*$
(for more details, see Balsara 1994, Mart\'{\i} \& M\"uller 1994 and references therein).
For a rarefaction wave, one can use the fact that the Riemann invariant 
\begin{equation}
J_{\pm}=\frac{1}{2}\,\ln\left(\frac{1+v}{1-v}\right) \pm \int \frac{c_s}{\rho}\,d\rho
\end{equation}
is constant through a rarefaction wave propagating to the left ($+$ sign) or to the right ($-$ sign).
The integral above can be evaluated analytically using the adiabatic flow condition $p/\rho^{\Gamma}=
const$. The equality of the Riemann invariants at 
the head and tail of the rarefaction wave and the fact that the entropy is constant throughout
the wave yield:
\begin{equation}
v_*=\frac{(1+v_0)A_{\pm}(p_*)-(1-v_0)}{(1+v_0)A_{\pm}(p_*)+(1-v_0)}\, , \quad
\rho_*/\rho_0=\left(p_*/p_0\right)^{1/\Gamma},
\label{velr}
\end{equation}
where
\begin{equation}
A_{\pm}(p_*)=\left[\frac{\left(\sqrt{\Gamma-1}-c_s(p_*)\right)
\left(\sqrt{\Gamma-1}+c_s(p_0)\right)}{\left(\sqrt{\Gamma-1}+c_s(p_*)\right)
\left(\sqrt{\Gamma-1}-c_s(p_0)\right)}\right]^{\pm2/\sqrt{\Gamma-1}}.
\end{equation}
For a shock wave, $\rho_*$ can be found by solving the Taub adiabatic condition (\cite{Taub}),
\begin{equation} 
(p_*-p_0)\left({h_0\rho_*^2}+h_*\rho_0^2\right) = h_0^2\rho_*^2-h_*^2\rho_0^2.
\end{equation}
$v_*$ is calculated using the pre- and post-shock flow velocities $(v)_{sh}$ and $(v_*)_{sh}$
in the shock's rest frame (\cite{Landau}): 
\begin{equation}
(v)_{sh}=\sqrt{\frac{(p_*-p_0)(\rho_*+e_*+p_0)}{(\rho_*+e_*-\rho_0-e_0)(\rho_0+e_0+p_*)}}\, , \quad
(v_*)_{sh}=\sqrt{\frac{(p_*-p_0)(\rho_0+e_0+p_*)}{(\rho_*+e_*-\rho_0-e_0)(\rho_*+e_*+p_0)}} \,. 
\label{vels}
\end{equation}
The first equation above and the pre-wave $v_0$ are used to determine the speed of the shock, needed 
to calculate the laboratory frame post-shock velocity $v_*$ from the second equation.

 The computation of the post-wave states is reduced now to solving the algebraic
equation $v_{*L} (p_*) = v_{*R} (p_*)$ for $p_*$, where $v_{*L}$ and $v_{*R}$ are determined
using equations (\ref{velr}) and (\ref{vels}).
Once $p_*$ is found, all other quantities $\rho_{*L}$, $\rho_{*R}$ and $v_*$ can be easily computed.
In our Riemann solver it is initially assumed that the discontinuity
decomposes into a shock and a rarefaction wave. A consistency check followed by a
feedback loop are used to ensure the initial choice of elementary waves 
(shock or rarefaction) is consistent with the values of the initial state pressure and the post-wave 
pressure after the last iteration:
if $p_* > p_{L/R}$ then the elementary wave is a shock, while if $p_* < p_{L/R}$
then it is a rarefaction wave moving toward left/right.
If these types of waves are not the same as those initially chosen,
then $p_*$ is used as the starting point for a new set of iterations, in which
the types of elementary waves are determined solely by 
$p_L$, $p_*$ and $p_R$. In most of our tests, the shock-rarefaction
approximation gives the correct solution for $p_*$, but we did find in a few
cases that, using this approximation, the iterated $p_{*,s-r}$ 
differs substantially from the true one. In these cases,
a few more iterations are needed to obtain the correct intermediate pressure $p_*$ after choosing the
elementary waves that are consistent with $p_L$, $p_{*,s-r}$ (calculated so far) and $p_R$.
Thus, our Riemann solver is exact in the sense that it yields self-consistent solutions.

 The second step in solving a Riemann problem, i.e$.$ in calculating its solution ${\bf U}(x,t)$ given the
initial left and right states and the position $x_0$ of the CD that separates them at $t=0$, is to determine
the location of the point of coordinate $x$ relative to the CD, the head(s) and tail(s) of the rarefaction
wave(s) and relative to the shock(s) at time $t$. This is done using the velocity at which
the CD, the rarefaction wave(s) boundaries and the shock(s) travel. 
The speed of the head of a rarefaction wave is the speed
of the sound in the initial left or right state. Similarly, the speed of the rarefaction wave tail is 
the sound speed in the corresponding post-wave state -- ($\rho_{*L},p_*,v_*$) or ($\rho_{*R},p_*,v_*$).
The CD travels at speed $v_*$ and the shock's speed is a by-product of the iteration procedure for
finding the post-wave states. In this way it is determined if after time $t$ the point $x$ is in one of the
yet unperturbed initial states, in a rarefaction fan or in one of the two uniform states around the CD.
If $x$ is not inside of a rarefaction fan, the solution ${\bf U}(x,t)$ is known: it is one
of the initial states or one of the post-wave states already calculated. If $x$ is in a rarefaction fan,
the first equation (\ref{velr}) (with $p$ and $v$ instead
of $p_*$ and $v_*$) and $(x-x_0)/\,t=(v \mp c_s)/(1 \mp v c_s)$ 
are used to determine by iteration $c_s$ and $v$ at ($x,t$),
after which $\rho$ and $p$ can be easily computed.
In problems with cylindrical or spherical symmetry,
similar to the work of \cite{sod77}, the geometrical effects 
are taken into account by an operator splitting technique: a Glimm step is followed
by a FD update using the geometrical terms in equations (\ref{dddt})--(\ref{dvdt}). 

 The Glimm method is implemented in relativistic hydrodynamic problems with non-uniform 
initial states. The fluid is approximated by a large number of cells of uniform states
${\bf u}_j^n \equiv (\rho,p,v)_j^n$,
centered at grid point $j$ and time-step $n$.
The Glimm method naturally calls for the use of a staggered computational mesh 
(see figure 1). 
Given two adjacent states ${\bf u}_j^n$ and ${\bf u}_{j+1}^n$ at time-step $n$,
the value of the approximate solution at time-step $n+1/2$ and position
$j+1/2$ is taken to be
the exact solution ${\bf U}(x,t)$ of the Riemann problem consisting of the left and
right states separated at time $n\Delta t$ by a fictitious CD located in the middle of the $(j,j+1)$ cell, 
evaluated at a randomly chosen point inside that cell: ${\bf u}_{j+1/2}^{n+1/2} = 
{\bf U}\left[(j+\xi_n)\Delta x,
(n+1/2)\Delta t \right]$, where $\xi_n$ is a random number in the interval [0,1].
The random number generator used in this work is based on the binary expansion of $n$:
\begin{equation}
\xi_{n}=\sum_{k=0}^{m}i_{k}2^{-(k+1)} \quad {\rm where} \quad  n=\sum_{k=0}^{m} i_{k}2^{k}\;\; (i_k=0,1).
\end{equation}
As for any random number generator, its quality can be assessed by determining
how fast the proportion of times that a generated number
is in a sub-interval $I$ of the interval [0,1] approaches the length of $I$, for any $I$. 
Colella (1982) compared this random sampling procedure with others commonly used and showed that 
it introduces the smallest errors associated with the sampling process characteristic to the Glimm method.
Moreover, these random numbers are chosen alternatively from
[0,0.5] and [0.5,1], so that a spurious propagation of a stationary CD is avoided.

In order to avoid the interaction of elementary waves generated
by Riemann problems in adjacent cells, the time-step must be chosen to satisfy 
$\Delta t < \Delta x \left(1+|v|c_s\right)/\left(|v|+c_s\right)$.
Note that this is not a condition for numerical stability. A stronger upper limit on $\Delta t$
can be imposed by requiring that the time-step is much smaller than the characteristic
time in which the wave speeds change, so that such changes are correctly treated by repeated sampling.
This restriction is obviously very important; nevertheless, all tests performed on the code based
on the Glimm method showed no correlation between the errors of the numerical solution and the
time-step size, as long as the computational grid contained more than 100 equal zones. In other
words, in all the tests considered, the maximum $\Delta t$ allowed by the above inequality is 
small enough to lead to a sufficiently large number of samplings in those regions of the fluid where
fast wave speed changes occur. The second criterion for the maximum $\Delta t$ is effective on 
coarser computational grids (less than 100 grid zones) where, if $\Delta t$ is set too large ($\sim \Delta x$),
the numerical solution may not be sufficiently smooth and the location of discontinuities may be 
inaccurate by several grid zones.

By construction, the Glimm method does
not require tracking shocks and CDs or a decrease of the grid cell size in
regions of the fluid with sharp gradients. The main advantages of the Glimm method are:
(1) it produces completely sharp shocks and CDs, thus allowing the use of
   uniform grids for treating discontinuities, 
(2) it is free of diffusion and dispersion, and 
(3) it is conservative on the average over time. 
It is not conservative over 
the spatial grid: in the tests considered below, 
the relative errors in the mass and in the energy 
of the fluid fluctuate around zero with
an amplitude of order 1\%, for a grid consisting of 100 equidistant zones.
As expected, the finer is the computational mesh, the lower is the amplitude of these fluctuations.
The drawback of the Glimm method is that
solving by iteration the equation $v_{*L}(p_*)=v_{*R}(p_*)$ for the post-wave pressure $p_*$
may lead to expensive runs in problems involving large evolution times.
One such iteration in the Riemann solver is $\sim 10$ times
more time consuming than a simple FD scheme. Usually 2 to 7 iterations are necessary in order to
determine $p_*$ with a good accuracy, depending on how close the left and right states are,
therefore the Riemann solver is 20 to 70 more computationally expensive than a FD scheme.
In addition, vectorization seems not to be possible 
with this method. Our approach to overcome this problem is to combine
the ability of the Glimm method to resolve discontinuities
with high-speed FD schemes, thus constructing a hybrid 
(or shock-patching) code.
We considered two FD schemes to solve equations (\ref{dddt})--(\ref{dvdt}) in the smooth parts of
the fluid: Lax and two-step Lax-Wendroff (\cite{recipes}). 
The codes that we developed to merge
these FD schemes with the Glimm method will be referred to as the Lax 
and LW codes, respectively.
This approach does not depend on the choice of the FD method
since they are only meant to be applied on regions free of discontinuities.
Besides their simplicity, another reason for choosing the Lax and Lax-Wendroff schemes 
is that they naturally work on the staggered grid
required by the Glimm method (see figure 1).

For the present work, in addition to the hybrid Lax and LW codes,
we also tested a code which
uses the Glimm method throughout the entire computational domain, the G code.
The hybrid codes are faster than the G code and
yield smoother profiles because of their numerical viscosity.
One important ingredient in the design of these hybrid
methods is the algorithm for the detection of discontinuities.
This sharp-features detector must be such that 
the smearing of large gradients
is prevented and, at the same time, the over-use of the Riemann solver
in reasonably smooth regions is avoided.
In practice, we found that the most robust detection algorithm
is based on the relative change of physical quantities of neighboring states;
that is $\epsilon \equiv |\phi_1 - \phi_2|/(\phi_1 
+ \phi_2)$, with $\phi = p$ or $\rho$.
Typically, the criterion to apply the Glimm method was that $\epsilon > 0.1$, but
values as large as 0.5 were used in problems involving large jumps across
discontinuities. Occasionally, the Glimm
method may be chosen instead of the FD scheme in a cell that does not contain a discontinuity,
if large gradients occur in that cell, as the discontinuity detector may consider
a large gradient to be a discontinuity.
The G code was used to calibrate the discontinuity detector and
to estimate the effects of the numerical viscosity inherent in the hybrid codes.

\section{Code Testing}

We have considered the following tests to evaluate 
the ability of the G, Lax and LW codes to simulate the propagation and
reflection of shocks and CDs:
(1) relativistic shock tube problems,
(2) shock heating of a cold fluid in planar and spherical geometry,
(3) cylindrical and spherical shock reflection and
(4) collisions of a cold relativistic shell 
with a stationary medium. 
For these tests, we calculate L1 norm errors
${\cal E}_1(T)=\sum_i \;[(r_{i+1/2})^{\alpha+1}-(r_{i-1/2})^{\alpha+1}] 
\; |T(r_i)-T_i|$ ($\alpha$ is the numerical coefficient of the geometrical
terms), mean relative errors 
${\bar {\cal E}}(T)=N^{-1}\sum_i^N |T(r_i)-T_i|/|T(r_i)|$,
and maximum relative errors
${\cal E}^{max}(T)= \max\,\lbrace |T(r_i)-T_i|/|T(r_i)|\rbrace$, 
with $T(r_i)$ the exact solution. The quality of a any of the three codes
is also assessed through the convergence rate $R$ of the numerical 
solution, defined as the limit of $d\ln {\cal E}_1/d\ln \Delta r$ when $\Delta r \rightarrow 0$.
We set the time-step close to the maximum value allowed by the FD 
stability criterion for relativistic flows ($\Delta t = \Delta r$),
which also satisfies the Glimm method condition for non-interaction
of waves from adjacent cells. 
However, the time-step was occasionally decreased 
if the occurrence of steep gradients would 
lead to an excessive evacuation of material. Such situations occur in the G code, near the
origin $r=0$ in problems with cylindrical and spherical symmetry, due to the geometrical terms,
and in the LW code in the very steep rarefaction fan that develops in the relativistic blast wave 
test described below.
The average CPU time per numerical time-step ($\Delta t/2$) and per grid cell, on a
Sun Sparc 20 Station, is $5 \div 6 \; \mu {\rm s}$ for the hybrid codes.
The CPU time of the hybrid codes changes little from one test to another, as long
as it is dominated by finite differencing computations. A substantial fraction of the computational time
could be used to solve Riemann problems if the numerical grid is made of less than 50 zones, in
which case the CPU time may vary from one test to the next, reflecting the
number of iterations done in the Riemann solver. The G code CPU time depends stronger on the problem that is
solved. These times (from runs done on the same machine) are given at the end of the following subsections
describing the numerical tests. 
They show that the hybrid codes are always faster than the G code.
The Riemann solver is used at discontinuities in all codes, therefore the higher speed of the hybrid codes
does not come from the way discontinuities are handled but from the fact that the Lax and LW codes 
do fewer computations than the G code in regions free of discontinuities. 
The Riemann solver requires only 1--2 iterations in those cells where the flow is smooth,
since the left and right states are close. For this reason,  
the hybrid codes are on average $5 \div 10$ times faster than the G code.
 
\subsection{Shock Tube Problems}

Relativistic shock tube problems 
consist of a discontinuity separating at $t=0$
two uniform static fluids.
As the fluid with larger pressure relaxes, a rarefaction wave
propagates into it and a shock sweeps up the lower pressure fluid. 
Two shock tube problems are considered:
Problem 1 has initial conditions $(p,\, \rho)_L = (13.3,\, 10)$,  
    $(p,\, \rho)_R = (0.66\times 10^{-6},\, 1)$, and
Problem 2 with initial states $(p,\, \rho)_L = (10^3,\, 1)$,  
    $(p,\, \rho)_R = (10^{-2},\, 1)$.
Subscripts $L$ and $R$ denote left and right states, respectively.
In both problems, the initial CD is placed at $x_0=0.5$,
and the adiabatic index is $\Gamma=5/3$. 
These initial conditions were chosen so that direct comparisons with the results 
obtained by \cite{hawley84} and \cite{marti96} for 
Problem 1 and \cite{norman86}, \cite{marti96} and \cite{falle96}
for Problem 2 are possible;
\cite{balsara94} also considered these two problems. 
Figure 2 shows profiles of $p$, $\rho$ and $v$ at $t=0.36$, 
generated by the three codes on a uniform mesh of $400$ cells, compared against the 
analytic solution.
An important feature to be resolved in this type of problem is
the characteristic thin and dense region between 
the shock and the CD.
The thickness of this region is $\Delta=4.12\times 10^{-2}$ in Problem 1
and $\Delta=9.50\times 10^{-3}$ in Problem 2.
In Problem 1, the Lorentz factor in the shocked fluid is $\gamma_{max}=1.43$
and that of the shock itself is $\gamma_{sh}=1.78$ .
The corresponding values for Problem 2 are
$\gamma_{max}=3.59$ and $\gamma_{sh}=6.18$ .

All codes give the correct position of the shock
and CD, up to one grid-spacing.
The constant states are well realized and discontinuities are infinitely sharp.
Note that the Lax code smears
the rarefaction head over more zones than does the LW code, while the G code
gives a perfectly sharp corner. We find that the rarefaction fans generated
with the G code are not completely smooth due to the randomness in the Glimm scheme 
(Sod 1978). Table 1 lists the L1 norm errors 
in mass, momentum and energy at $t=0.36$, for $\Delta x=1/400$ in the
first problem and $\Delta x=1/800$ in the second problem. 
Also given in this table are the maximum and average relative errors 
in mass and energy conservation reached during run, as well as 
the convergence rates ($\cal R$) of the solutions computed using grids consisting of
50, 100, 200, 400, 800, 1600 and 3200 equal zones.
The convergence rates show that the Glimm method is first order accurate,
in agreement with the analytic study by \cite{liu77}. Liu showed
that the convergence rate of the Glimm scheme is at most 1,
depending on how well the sequence of random numbers $\xi_n$ is distributed in the
interval [0,1].  The effect of the randomness involved 
in the updating of physical quantities
can be seen in the solutions obtained with the G code. 
In most cases, they have the largest
maximum and average relative errors in mass and energy conservation.
These shock tube tests also demonstrate 
that combining the Glimm method with the Lax scheme
yields a code that is less than first order 
accurate although, independently,  both schemes are first order convergent.
The CPU time per timestep and per 
grid zone for the G code is $21\; \mu {\rm s}$ in Problem 1 and  $26\; \mu {\rm s}$ in Problem 2.
For an equal computer time, the Lax code can
work on a twice finer grid, producing solutions with comparable accuracy
to those from the G code.
Combining the Glimm method with the second-order Lax-Wendroff scheme leads to a first
order accurate LW code because the errors at discontinuities dominate the overall
L1 norm errors. The LW code yields more accurate solutions than does the Lax code and has
a higher convergence rate.

\subsection{Shock Heating Test}

The problem of a 1D cold fluid hitting a wall and generating a shock that 
propagates into the incoming fluid and heats it, has been considered
by many authors: \cite{centrella84}, \cite{hawley84}, \cite{norman86}, 
\cite{marti96} and \cite{falle96}. Behind the shock, 
the fluid is at rest and hot ($p\gg \rho$), provided
that the impacting fluid is relativistic. The shock jump conditions 
(\cite{blandford76}) yield
the post-shock pressure $p_2$, density $\rho_2$ and shock velocity $v_{sh}$:
\begin{equation}
p_2=\rho_1 \left( \gamma_1-1 \right) \left(\gamma_1 \Gamma_2+1 \right)  \qquad
\rho_2=\rho_1 \frac{\gamma_1 \Gamma_2 +1}{\Gamma_2 -1} \qquad
v_{sh}= - \left( \Gamma_2 -1 \right) \frac{\gamma_1 v_1}{\gamma_1 +1}\, ,
\label{wall}
\end{equation}
where the subscript $1$ refers to pre-shock quantities and $\Gamma_2$ is the
adiabatic index of the shocked fluid (taken to be 4/3). 
Pressure, density and flow velocity profiles are uniform
ahead of and behind the shock. Since all codes use 
the Glimm method in the fluid cell that contains the shock, there is no
substantial difference in the results given by any of the three codes. 
This planar shock heating problem is basically used 
to test the ability of the codes
to calculate the maximum and average error in the compression factor
$\eta=\rho_2/\rho_1$ and to
see how this error changes as the impacting fluid becomes more and more
relativistic.
It is important to notice that we find that the errors in this problem 
are determined only by the Riemann solver's accuracy in calculating 
the post-shock pressure in the shock's cell. Depending on the limit at
which the Riemann solver converges within double precision, the maximum
and average errors in $\eta$ are between $10^{-10}$ and $10^{-11}$, many 
orders smaller than the values quoted by other authors. 

The shock heating problem in spherical geometry is
more difficult and interesting since it tests
the Glimm method under an operator splitting approach to handle the geometrical terms. 
This problem consists of a cold fluid entering a sphere of radius unity 
at constant velocity $v_0$ and was used as a test by \cite{romero96} and \cite{marti97}.
It has an analytic solution if the fluid 
is initially homogeneous $\rho(0,r)=\rho_0$ and if, at any time, particles 
in the yet un-shocked fluid flow independently, with constant velocity
$v(t,r)=v_0$ (i.e. the pressure of the cold gas is negligible and does not
decelerate the relativistic inflow).
From mass conservation, the density of the un-shocked fluid is 
$\rho_1(t,r)=\left(1+\, |v_0|t/r \right)^2 \rho_0$. The post-shock 
density $\rho_2$ is determined by the density of the un-shocked fluid 
at the shock position, therefore (from eq.[\ref{wall}]) $\rho_2= 
\rho_0 \left[\left(\Gamma_2 \gamma_0+1\right)/\left(\Gamma_2-1
\right)\right]^3\gamma_0^{-2}$. The post-shock density $\rho_2$ is position
and time independent.

Figure 3 shows $p$, $\rho$ and $|v|$ at $t=1.90$
for a 400 zone grid and initial conditions
$\rho_0=1$, $p_0 = 10^{-10}$ and $v_0=0.99999$ ($\gamma_0 \simeq 224$).
Notice that the Lax code (see figure 3, upper graph) 
produces smoother profiles near the origin, and thus it yields smaller errors
than the LW and G codes do (figure 3, middle and lower graphs, respectively).
This is due to the high numerical viscosity of the Lax scheme. 
The LW and G codes solutions are dominated by errors near the origin 
because of the coordinate singularity (for regularization, see for instance \cite{putten}). 
No attempt has been made to reduce these errors since, as we mentioned before, our main motivation
for developing a hydrodynamic code is to simulate the propagation of the ultra-relativistic shocks likely to
be present in GRBs, in which case the computational region of interest 
is considerably away from the origin.
Table 2 lists the maximum and average relative errors in the ratio
$\eta=\rho_2/\rho_0$ at $t=1.90$, on a 400 equidistant zone grid, for different inflow velocities.
The average relative errors are in most cases less than 1\% and show no correlation
with $\gamma_0$.
These errors can be compared with those published by 
\cite{romero96} who found that, in the more relativistic regime,
the maximum and average relative errors in $\eta$
are 14\% and 2.2\%, respectively, independently of the inflow velocity.
We note that only the G code gives maximum errors larger than 14\%, and 
that the Lax code maximum errors are smaller by a factor $\sim 10$.
The errors in the numerical solution obtained by \cite{romero96} are due
to a dip in the post-shock density near the origin, which
\cite{noh87} showed that is caused by the fact that
the constant pressure gas is hotter near the origin -- an effect
due to the artificial viscosity technique they used. 
For $\gamma_0=224$, the average convergence rates of
the numerical solutions are 1.00, 1.00 and 0.86 for the Lax, LW and G code, respectively.
The CPU time per timestep and per
grid zone for the G code increases from $58\; \mu {\rm s}$ for $\gamma_0=2.3$ up to
$190\; \mu {\rm s}$ for $\gamma_0=2236$.

\subsection{Cylindrical and Spherical Shock Reflection}

This problem consists of a CD 
separating two uniform states. Initially,
the fluid is at rest and the pressure outside the CD is larger
than inside, generating a shock that propagates 
radially inward and is reflected in the origin.
The left and right states are:
$p_L=\rho_L=1.0$ and $p_R=\rho_R=4.0$, thus the fluid is hot
($\Gamma=4/3$). The CD is initially located at $r_0=0.5$.
The same problem, but only in cylindrical symmetry and 
in Newtonian hydrodynamics, was considered by \cite{sod77}.
Since the G code cannot miss any discontinuity, 
we used this code to test how well the
hybrid codes detect and propagate discontinuities and 
simulate their interaction.
We found a satisfactory agreement among the solutions furnished by all codes, 
not only in the position and strength of all
discontinuities but also in the profiles in smooth fluid regions.

Figure 4 shows the density $D$, internal energy density $E_{int}=\gamma^2 h-p-\gamma D$
and velocity $v$ in the lab-frame, 
when the inward shock (innermost discontinuity in figures) is close to origin (thin dotted curves),
after it is reflected
(thick dotted curves), before the outward reflected shock interacts
with the CD (outermost discontinuity) between the two fluids (thin solid curves)
and after their collision (thin and thick dashed curves). In the cylindrical symmetry case,
the inward shock is reflected at $t=0.75$ and hits the CD at $t=1.16$ .  
The corresponding times in the spherical case are are 0.71 and 1.15 .
In cylindrical symmetry, the outward shock interacts with a 
CD moving in opposite direction, while in the spherical case,
the outgoing shock interacts with a CD
slowly moving in the same direction, as it can be seen in the lower graphs. 
There is a substantial discrepancy in the shock's speed between our solution and
that obtained by \cite{sod77} (shocks propagate faster in Newtonian hydrodynamics).
We list in Table 3 the errors of the solutions generated by each code on
a mesh made of 400 equal zones, at $t=1.0$ . Also in this table, we report
the convergence rate $\cal R$ of the mass, momentum
and energy solutions obtained from simulations with 
100, 200, 400 and 800 zones. Since there is no analytic solution to
this problem, the solutions calculated using 3200 zones were used as exact.
Note that solutions furnished by all codes have comparable convergence rates and that
these rates are closer to 1 in the cylindrical symmetric case. This suggests
that the severity of this test is partially determined by the ``strength" 
of the geometrical terms ($1/r$ terms) in equations (\ref{dddt})--(\ref{dvdt}).
The CPU time per timestep and per
grid zone for the G code is $58\; \mu {\rm s}$ in both problems.

\subsection{Relativistic Blast Wave}

Finally, we consider the interaction between a cold,
relativistically expanding
shell (initial Lorentz factor $\gamma_0$) with a cold, less dense stationary medium.
As the collision evolves, the shell is decelerated,
and the kinetic energy is gradually transformed into internal energy. At the same time, a shock
propagates inwards in the relativistic shell (reverse shock),
with a forward shock (or blast wave) sweeping up the external medium.
The thermal Lorentz factor of shocked fluid is comparable with
its bulk Lorentz factor, which can be roughly approximated by the
Lorentz factor $\gamma_0$ of the yet un-shocked fluid.
Therefore, the internal energy of
the shocked external medium is comparable with the shell's initial kinetic energy 
when the shell has swept up a fraction
$\gamma_0^{-1}$ of its mass. This occurs at the deceleration radius 
$r_{dec}= (3\,M/4\pi \gamma_0 n_0 m_p c^2)^{1/3}$, with $M$ 
the shell's mass, $n_0$
the number density of the stationary medium and $m_p$ the proton's mass.
Deceleration effects are 
important when the expanding shell reaches radial coordinates of order $r_{dec}$.
In this particular problem, we consider a shell of mass $M \sim 1.2 \times
10^{29}$ g, moving at
$\gamma_0=10$ into a medium with number density $n_0=1$ particle/${\rm cm}^3$.
Therefore,
$r_{dec} \sim 1.2 \times 10^{17}$ cm. The time for this shell
to reach $r_{dec}$ is $t_{dec}= r_{dec}/c \sim 4.0 \times 10^6$ s (deceleration
time-scale).

We set the initial position of the shell
at $r_0=0.4\,r_{dec}$ and consider that the external medium becomes effective 
only for $r > r_0$ (deceleration effects are negligible before $t_0=r_0/c=0.4\,t_{dec}$
since a fraction of only 6.4\% of the external medium mass within $r_{dec}$ has
been swept up so far).
The shell's lab-frame thickness at this position is set to a
fraction 1/100 of $r_0$; consequently,
at $t_0$ the shell lab-frame number density is
$N_0 \sim 5.2 \times 10^3\, {\rm cm}^{-3}$. 
Figure 5 shows the co-moving ($n$) and lab-frame ($N$)
number densities
inside the shocked structure, at $t=$ 1.2, 1.4, 1.6, 1.8 and 2.0
$t_{dec}$ (before $t=1.2\, t_{dec}$ the reverse shock crosses the 
inner shell and leads to the formation of a
steep rarefaction fan behind the CD).

In order to test the convergence
of numerical solutions, we used the solution generated on a mesh consisting
initially of 800 equidistant zones 
as an exact one. During the collision, the shocked structure expands;
at $t_{end}=2.0\,t_{dec}$, it is $\sim 25$ times thicker than at $t_0$. 
As the mesh is refined, the location of the forward shock
does not change, but the position of the CD between 
the two shells at $t_{end}$ shifts within a region $\sim 8 \times
10^{-3}\,r_{dec}$ thick. However, an uncertainty in the location of the CD
of 0.4\% of the distance traveled is not surprising when a
relativistic shell is evolved over time-scales 400 times larger than its
crossing time. To avoid over-estimating the errors due to the small off-set
in the position of the CD, we make a separate analysis of solutions
for each shell, and we align the CDs of the inner
shell profiles before calculating the L1 norm errors of their solutions.
Table 4 lists these errors in each shell solution at
$t_{end}$, for the Lax code. The grid-spacing $\Delta r$ is given in deceleration radii.
The G code can be used
on the coarser grids, but it becomes too expensive on the finer ones. 
The LW code requires
an adaptive time-step to avoid the un-physical pressures that occur when
a steep rarefaction wave develops in the inner shell, 
between $t \simeq 0.9\, t_{dec}$ and $t \simeq 1.2 \, t_{dec}$. 
This substantially decreases the speed of the LW code and leads to
errors comparable with those in the Lax code's solutions. Moreover, in this
smooth region with strong gradients, the LW code develops
a sequence of discontinuous jumps with the appearance of a staircase, much
like the effect described by \cite{woodward84} in connection with the use 
of the linear hybridization approach in flows with large gradients.

\section{Conclusion}

We have presented a numerical approach to solve 1D hydrodynamic problems,
suitable for calculating the long term evolution of ultra-relativistic shocks.
Our scheme uses
(1) the Glimm method, including an exact Riemann solver,
(2) an operator splitting technique to account for geometrical effects, and
(3) FD methods (Lax or Lax-Wendroff) in smooth regions of the computational
domain.
This shock patching approach takes advantage 
of the accuracy of the Glimm method to resolve
discontinuities and the computational speed of the FD methods.
Comparison of our results with those published by other authors shows
that the hybrid codes here developed are competitive and, possibly, faster.
This last feature makes them particularly suitable for problems requiring long
term evolution such as modeling the interaction of a cold shell with an external
medium at large Lorentz factors ($\gamma > 100$).

We thank Dr. P. M\'esz\'aros for helpful comments and suggestions on this work.
This research has been supported in part through NASA NAG5-2362 and NAG5-2857
to LW and AP, and through NSF-PHY 93-09834, NSF-PHY 93-57219 (NYI) to PL.


\clearpage

\input psfig

\begin{figure}
\centerline{\psfig{figure=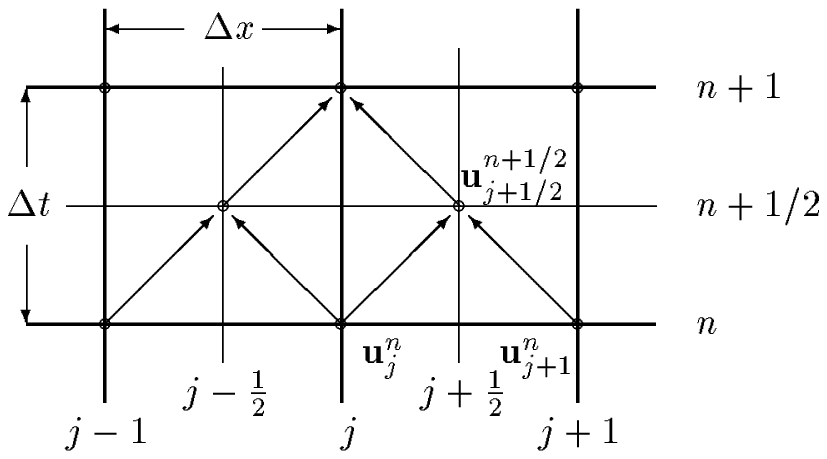}}
\vspace{-3 in}
\figcaption{Staggered mesh required by the Glimm method}
\end{figure}

\begin{figure}
\centerline{\psfig{figure=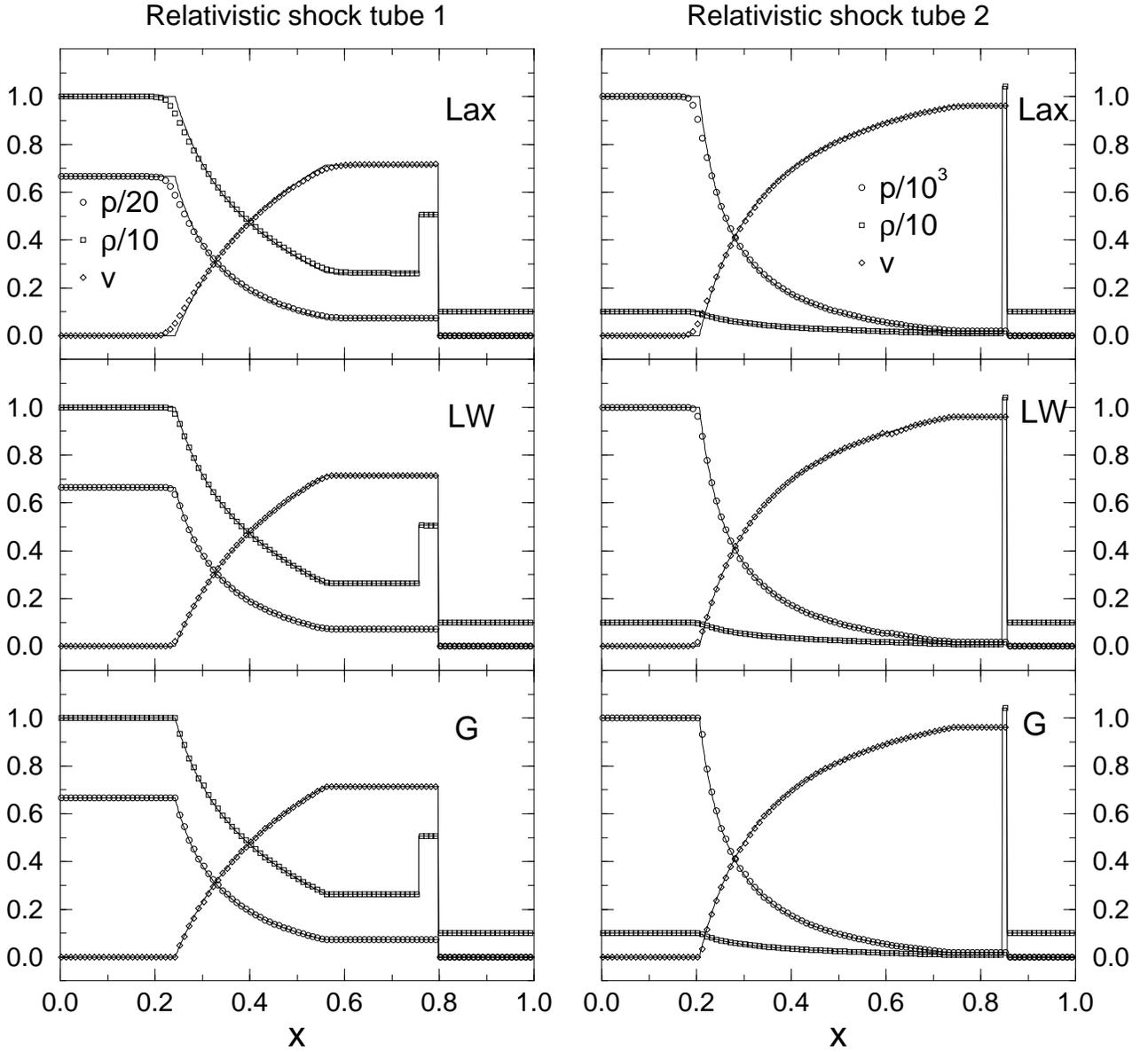}}
\vspace{1 in}
\figcaption{Solutions to the relativistic shock tube problems 
with 400 grid zones, at $t=0.36$. Left graphs for Problem 1 and
right graphs for Problem 2 (see text). 
First row: Lax code, second row: LW code, third row: G code.
The exact solution is represented with a solid line.}
\end{figure}

\begin{figure}
\centerline{\psfig{figure=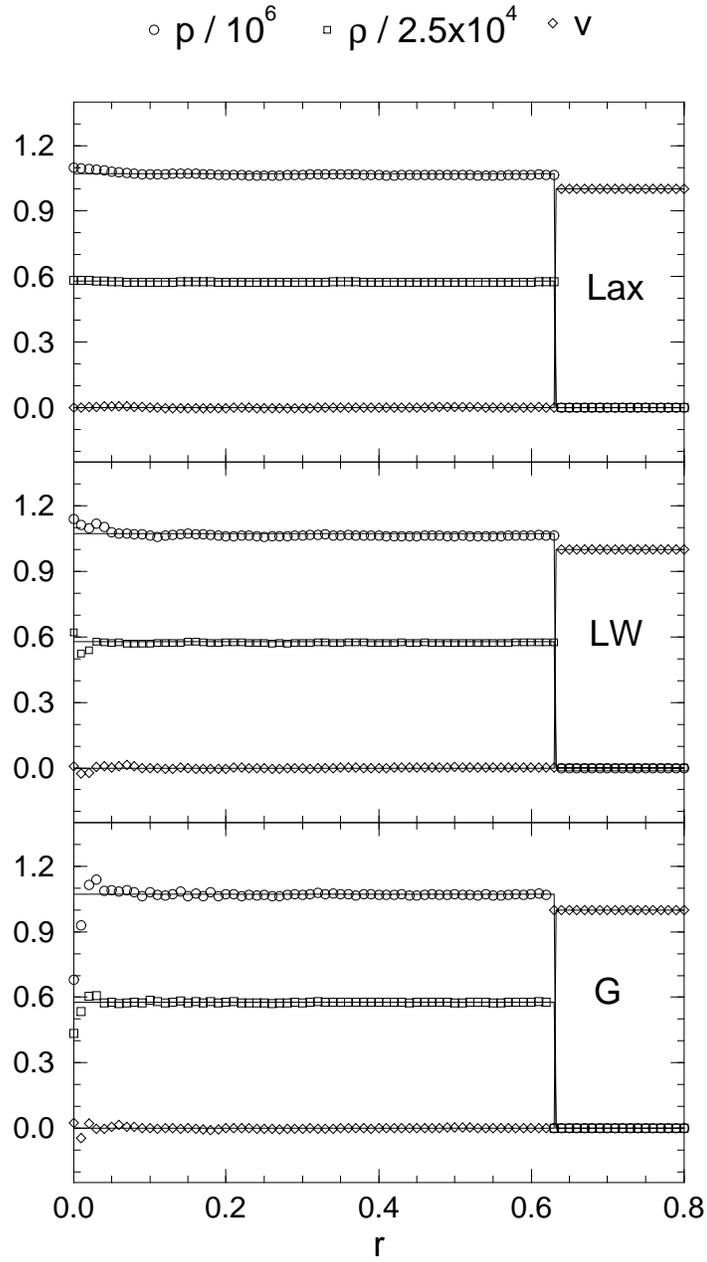}}
\vspace{1 in}
\figcaption{Solutions at $t=1.90$ for the 
spherical shock heating problem with inflow velocity $|v_0|=0.99999$
($\gamma_0 \simeq 224$), on a mesh of 400 zones, at $t=1.90$.
Solid curves show the analytic solution.
Upper graph: Lax code, middle graph: LW code, lower graph:
G code.}
\end{figure}

\begin{figure}
\centerline{\psfig{figure=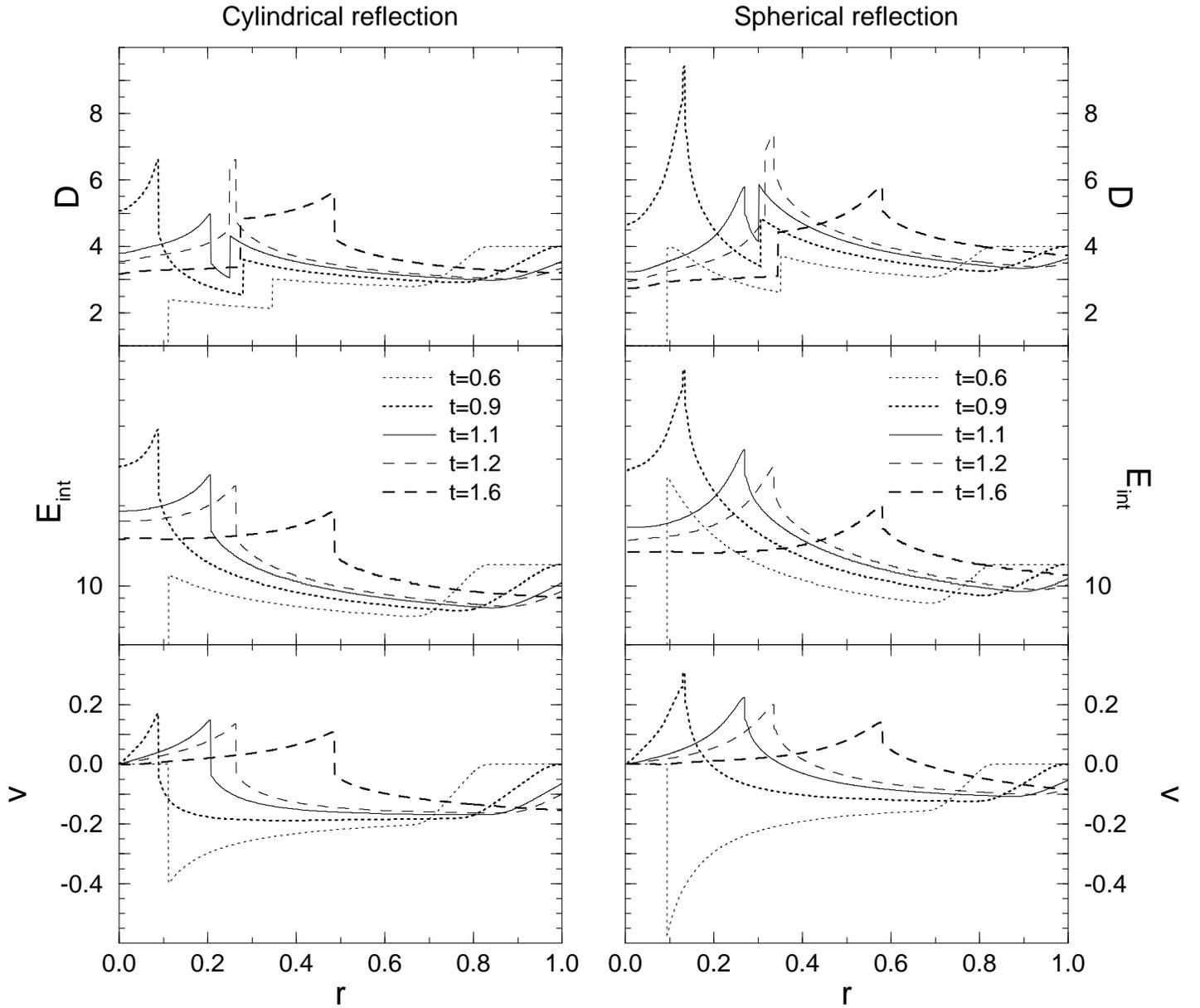}}
\vspace{1 in}
\figcaption{Cylindrical (left graphs) and spherical (right graphs) shock
reflection solutions using the Lax code with a 1000 zone mesh.
Note the larger
lab-frame density $D$, internal energy density $E_{int}$ (in logarithmic scale) and flow velocity $v$
near the origin in the spherical symmetric case. Negative velocity corresponds to a flow toward the
origin, a positive velocity indicates an outward fluid motion.
Profiles correspond to the times indicated in the legend.}
\end{figure}

\begin{figure}
\centerline{\psfig{figure=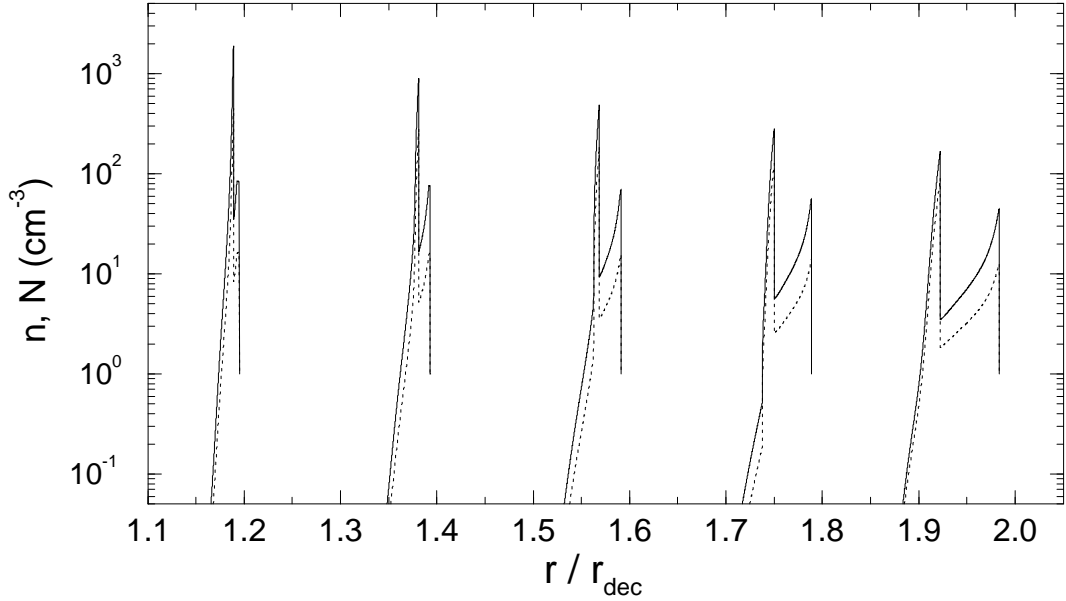}}
\vspace{1 in}
\figcaption{Relativistic ($\gamma_0=10$) blast wave: co-moving (dotted curves) and
lab-frame (solid
curves) number density at various times during the collision. From left to right, 
$t$= 1.2, 1.4, 1.6, 1.8 and 2.0 $t_{dec}$. Each profile shows the same structure from left to
right: inner shell (initially moving fluid) -- CD -- outer shell (shocked external fluid)
-- blast wave. The flow Lorentz factor can be estimated from the difference
between the co-moving and lab densities. Note the second reverse shock in the inner shell,
in the $t=1.6\,t_{dec}$ and $t=1.8\,t_{dec}$ profiles.}
\end{figure}

\clearpage

\begin{table}
\begin{center}

TABLE 1 \\ {\sc L1 norm errors ${\cal E}_1$,
maximum ${\cal C}^{max}$ and average $\bar{\cal C}$ relative errors (\%)
in mass (M) and energy (E) conservation during runs,
for shock tube problems} \\[4ex]

\begin{tabular}{cccccccccc} \hline \hline
\rule[-4mm]{0mm}{10mm} Problem & Code & 
${\cal E}_1$(D) & ${\cal E}_1$(S) & ${\cal E}_1$(E) &
${\cal C}^{max}_M$  &  $\bar{\cal C}_M$  &
${\cal C}^{max}_E$  & $\bar{\cal C}_E$  &
$\cal R$ \\ \hline \hline
\rule[-2mm]{0mm}{6mm} 1 & Lax & 7.21 E-2 & 1.33 E-1 & 1.34 E-1 &
                      0.31 & 0.12 & 0.50 & 0.24 & 0.79  \\ \hline
\rule[-2mm]{0mm}{6mm} 1 & LW  & 2.40 E-2 & 4.29 E-2 & 4.83 E-2 &
                      0.54 & 0.30 & 0.53 & 0.28 & 1.06  \\ \hline 
\rule[-2mm]{0mm}{6mm} 1 & G   & 3.09 E-2 & 5.00 E-2 & 3.92 E-2 &
                      0.89 & 0.50 & 0.92 & 0.55 & 1.02  \\ \hline \hline
\rule[-2mm]{0mm}{6mm} 2 & Lax & 1.30 E-3 & 3.38 E-0 & 4.30 E-0 &
                      4.72 & 1.61 & 0.22 & 0.13 & 0.64  \\ \hline
\rule[-2mm]{0mm}{6mm} 2 & LW  & 8.89 E-4 & 6.81 E-1 & 8.66 E-1 &
                      4.74 & 1.61 & 0.15 & 0.07 & 0.88  \\ \hline 
\rule[-2mm]{0mm}{6mm} 2 & G   & 3.69 E-4 & 3.33 E-1 & 5.06 E-1 &
                      4.78 & 1.60 & 0.33 & 0.19 & 0.97  \\ \hline \hline
\end{tabular}

\end{center}
\end{table}


\begin{table}
\begin{center}

TABLE 2 \\ {\sc Maximum ${\cal E}^{max}$  and average $\bar{\cal E}$ relative errors
(\%) in ratio $\eta=\rho_2/\rho_0$, in the spherical shock heating test} \\ [4ex]

\begin{tabular}{cccccccc} \hline \hline
\rule[-4mm]{0mm}{10mm} $1 - |v_0|$ & $\gamma_0$ & ${\cal E}_G^{max}$  &
${\cal E}_{Lax}^{max}$  & ${\cal E}_{LW}^{max}$  &
$\bar{\cal E}_G$  &
$\bar{\cal E}_{Lax}$  & $\bar{\cal E}_{LW}$  \\
\hline \hline
\rule[-2mm]{0mm}{6mm}  $10^{-1}$ &  2.3 & -22.2 & -2.57 & -10.1 &
                                           2.56 & 0.84 & 0.95 \\
\hline
\rule[-2mm]{0mm}{6mm}  $10^{-3}$ &  22  &  35.2 & -1.16 & -13.3 &
                                           0.93 & 0.47 & 1.01 \\
\hline
\rule[-2mm]{0mm}{6mm}  $10^{-5}$ &  224 &  10.4 & -1.07 & -10.1 &
                                           0.72 & 0.61 & 1.08  \\
\hline
\rule[-2mm]{0mm}{6mm}  $10^{-7}$ & 2236 &  11.9 &  1.52 & -10.1 &
                                           0.97 & 0.35 & 0.57 \\
\hline \hline
\end{tabular}

\end{center}
\end{table}

\clearpage

\begin{table}
\begin{center}

TABLE 3 \\ {\sc  L1 norm errors ${\cal E}_1$ in the
the cylindrical (left part of the table) and spherical (right part) shock
reflection problem} \\ [4ex]

\begin{tabular}{ccccccccc} \hline \hline
\rule[-4mm]{0mm}{10mm} $\cal R$ & ${\cal E}_1$(D) 
& ${\cal E}_1$(S) & ${\cal E}_1$(E) & Code &
            ${\cal E}_1$(D) & ${\cal E}_1$(S) & ${\cal E}_1$(E) & $\cal R$  \\ \hline \hline
\rule[-2mm]{0mm}{6mm} 0.96 & 1.66 E-2 & 3.79 E-2 & 6.88 E-2 & Lax &
                  2.14 E-2 & 3.28 E-2 & 8.43 E-2 & 0.65 \\ \hline
\rule[-2mm]{0mm}{6mm} 0.96 & 2.06 E-2 & 4.92 E-2 & 7.81 E-2 & LW &
                  1.27 E-2 & 2.53 E-2 & 4.38 E-2 & 0.84 \\ \hline
\rule[-2mm]{0mm}{6mm} 0.91 & 1.34 E-2 & 2.92 E-2 & 5.31 E-2 & G &
                  9.45 E-3 & 2.03 E-2 & 3.55 E-2 & 0.70 \\ \hline  \hline
\end{tabular}

\end{center}
\end{table}


\begin{table}
\begin{center}

TABLE 4 \\ {\sc L1 norm errors ${\cal E}_1$ in the inner shell (left part of the table) and
outer shell (right part) solutions for the relativistic blast wave problem} \\ [4ex]

\begin{tabular}{ccccccc} \hline \hline
\rule[-4mm]{0mm}{10mm} ${\cal E}_1$(D) & ${\cal E}_1$(S) & ${\cal E}_1$(E) & $\Delta r$ &
                       ${\cal E}_1$(D) & ${\cal E}_1$(S) & ${\cal E}_1$(E) \\ \hline \hline
\rule[-2mm]{0mm}{6mm} 7.13 E-3 & 1.16 E-2 & 6.53 E-3 & $1.6 \times 10^{-4}$ & 
                      5.73 E-4 & 1.82 E-2 & 1.78 E-2 \\ \hline
\rule[-2mm]{0mm}{6mm} 3.72 E-3 & 5.92 E-3 & 3.29 E-3 & $  8 \times 10^{-5}$ & 
                      2.11 E-4 & 6.79 E-3 & 6.66 E-3 \\ \hline
\rule[-2mm]{0mm}{6mm} 1.52 E-3 & 2.51 E-3 & 1.46 E-3 & $  4 \times 10^{-5}$ & 
                      1.81 E-4 & 3.21 E-3 & 3.10 E-3 \\ \hline
\rule[-2mm]{0mm}{6mm} 8.98 E-4 & 1.42 E-3 & 7.83 E-4 & $  2 \times 10^{-5}$ & 
                      9.91 E-5 & 2.89 E-3 & 2.83 E-3 \\ \hline \hline
\rule[-3mm]{0mm}{8mm} 1.03 & 1.03 & 1.04 & $\cal R$ & 0.65 & 0.78 & 0.90 \\
\hline \hline
\end{tabular}

\end{center}
\end{table}

\end{document}